\documentclass[aps,pre,amssymb]{revtex4-2}
\usepackage{graphicx}
\usepackage{color}

\begin{document}

\title{Transient anomalous diffusion in heterogeneous media with stochastic resetting}

\author{M. K. Lenzi$^{1}$, E. K. Lenzi$^{2}$, L. M. S. Guilherme$^{2}$, L. R. Evangelista$^{3,4}$, H. V. Ribeiro$^{4}$}
\affiliation{$^1$Departamento de Engenharia Química, Universidade Federal do Paraná, Curitiba, Paran\'a, Brazil.\\
$^{2}$Departamento de F\'isica, Universidade Estadual de Ponta Grossa - Ponta Grossa, Paran\'a, Brazil\\
$^{3}$Departamento de F\'{\i}sica, Universidade Tecnol\'ogica Federal do Paran\'a,  Apucarana, Paran\'a, Brazil\\
$^{4}$Departamento de F\'isica, Universidade Estadual de Maring\'a - Maring\'a, Paran\'a, Brazil}

\begin{abstract}
We investigate a diffusion process in heterogeneous media where particles stochastically reset to their initial positions at a constant rate. The heterogeneous media is modeled using a spatial-dependent diffusion coefficient with a power-law dependence on particles' positions. We use the Green function approach to obtain exact solutions for the probability distribution of particles' positions and the mean square displacement. These results are further compared and agree with numerical simulations of a Langevin equation. We also study the first-passage time problem associated with this diffusion process and obtain an exact expression for the mean first-passage time. Our findings show that this system exhibits non-Gaussian distributions, transient anomalous diffusion (sub- or superdiffusion) and stationary states that simultaneously depend on the media heterogeneity and the resetting rate. We further demonstrate that the media heterogeneity non-trivially affect the mean first-passage time, yielding an optimal resetting rate for which this quantity displays a minimum.
\end{abstract}

% \pacs{05.40.-a,05.40.Jc,05.10.Gg}

\maketitle

\section{Introduction}

After the experiments conducted by R. Brown and the pioneer works of A. Einstein~\cite{einstein}, P. Langevin~\cite{langevin}, M. Smoluchowski~\cite{von1906}, and K. Pearson~\cite{KP}, diffusion and its connection with stochastic processes have become crucial concepts for several fields of science~\cite{Book2018,ionmotion,lippincott2001studying,albinali2016anomalous,RAZMINIA2020105068,metzler1994fractional}. In addition to several applications and formal developments, diffusion is considered an essential phenomenon in living cells, as it plays an important role in cellular processes such as nuclear organization, division, differentiation, and migration~\cite{kinkhabwala2010spatial,capoulade2011quantitative,Living1,Living2}. Depending on the media and the interactions among system components, diffusion processes are classified as usual or anomalous. A fingerprint of usual diffusion is the proportional increase of the mean square displacement with time, that is, $\langle \left(x-\langle x \rangle\right)^2\rangle \sim t$. Anomalous diffusion in its turn is associated with a power-dependence for the mean square displacement, that is, $\langle \left(x-\langle x \rangle\right)^2\rangle\sim t^{\gamma}$, where $\gamma<1$ and $\gamma>1$ correspond to sub- and superdiffusion, respectively. Diffusion often emerges combined with different phenomena such as stochastic resetting~\cite{evans2011diffusion,evans2014diffusion}, a process in which particles are stochastically repositioned to their initial positions at a constant rate. Examples of systems with stochastic resetting include the production of proteins by ribosomes~\cite{nagar2011translation}, visual working memory in humans~\cite{balaban2017neural}, protein identification in DNA~\cite{Reuveni4391}, and animal foraging~\cite{bartumeus2009optimal}. 

Motivated by this myriad of possible applications, several works have systematically investigated the combination of diffusion with stochastic resetting~\cite{bhat2016stochastic,Evans_2013,sandev2021diffusion,pal2016diffusion,pal2015diffusion,evans2020stochastic,doi:10.1063/5.0010549,ray2021resetting,shkilev2017continuous,dos2019fractional,kusmierz2019subdiffusive,PhysRevE.101.022135}. Despite this increasing interest in studying diffusion with stochastic resetting, much less attention has been paid towards considering the effect of heterogeneous media~\cite{ray2020space}. This is an important aspect of diffusive processes in several systems such as diffusion on fractals~\cite{PhysRevLett.54.455,PhysRevA.32.3073}, correlated random velocity fields~\cite{PhysRevE.60.5528}, atmosphere~\cite{richardson1926atmospheric}, and under turbulence~\cite{book1371194,book1371197}. 

Here we help to fill this gap by investigating a diffusive process in heterogeneous media with stochastic resetting. To account for media heterogeneity, we consider a diffusion coefficient with a power-law dependence on particles’ positions~\cite{PhysRevLett.54.455,PhysRevA.32.3073,PhysRevE.60.5528,richardson1926atmospheric,PhysRevLett.88.094501,book1371194,book1371197,book321195,PhysRevLett.102.045901}, that is, $D(x) \sim |x|^{-\eta}$, where $\eta>-1$ is a parameter. 
This spatial dependence for the diffusion coefficient emerges in several situations and is often associated with solutions in terms of a stretched exponential distributions. Such distributions are typical in different systems, including diffusion on fractals~\cite{PhysRevA.32.3073,PhysRevLett.54.455}, turbulence~\cite{richardson1926atmospheric,PhysRevLett.88.094501}, diffusion and reaction on fractals~\cite{book321195}, solute transport in fractal
porous media~\cite{SU2005852}, atom deposition in a porous substrate~\cite{PhysRevLett.102.045901}. This particular dependence for the diffusion coefficient connects our investigation with these systems or other ones having similar structures. Furthermore, the parameters $r$ and $\eta$ allow us to investigate the interplay between stochastic resetting and media heterogeneity on the properties of this system, revealing a rich diffusive scenario marked by usual non-Gaussian distributions, different diffusive regimes (usual, sub-, and superdiffusion), and stationary states. We obtain these results by using the Green function approach which in turn allowed us to find exact solutions to the mean square displacement, probability distribution of particles' positions, and first-passage time. We further compare some of these exact results with simulations of Langevin-like equations. Among other results, we show that the plateau characterizing the mean square displacement in the stationary state depends not only on the resetting rate ($r$) but also on the media heterogeneity ($\eta$). Similarly, we find that the media heterogeneity affects the mean first-passage time in a non-trivial manner.

The rest of this manuscript is organized as follows. Section II defines the diffusion equation, presents the approach used to find its solution, and describes the results about the probability distribution of particles' positions and the mean square displacement. Section III describes the results associated with the first-passage time problem. Finally, Section IV concludes this works with an overview of our main findings.

\section{Diffusion on heterogeneous media with stochastic resetting}

We start our analysis by considering a system subjected to the following diffusion equation
\begin{eqnarray}
\label{resetting1}
\frac{\partial}{\partial t}\rho(x,t)=\frac{\partial}{\partial x}\left\{ D(x)\frac{\partial}{\partial x} \rho(x,t)\right\}-r\left[\rho(x,t)-\delta(x-x')\right]\,,
\end{eqnarray}
where $\rho(x,t)$ represents the probability distribution of finding a particle around position $x$ at time $t$, $r$ is the rate under which particles stochastically reset their positions to $x'$, $D(x) = D|x|^{-\eta}$ is the position-dependent diffusion coefficient, $\eta>-1$ is a parameter associated with the media heterogeneity, and $D$ is a constant corresponding to the usual diffusion coefficient for $\eta=0$. It is worth noting that we can relax the constraint on the values of $\eta$ to $\eta>-2$ when considering only the results associated with the probability distribution $\rho(x,t)$. However, and as we shall see in the next section, the condition $\eta>-1$ is necessary to find a non-diverging fist-passage time. We further consider the boundary conditions $\rho(\pm \infty,t) = 0$ and the initial condition $\rho(x,0)=\varphi(x)$. This equation thus represents a diffusive process with stochastic resetting occurring on spatially heterogeneous media. As we shall demonstrate, depending on the values $\eta$, this system displays transient sub-diffusion or superdiffusion before approaching a stationary state that is also dependent on $\eta$. It

To solve Eq.~(\ref{resetting1}), we first define $\rho(x,t) = e^{-rt}\bar{\rho}(x,t)$ and apply this change of variable to Eq.~(\ref{resetting1}), yielding
\begin{eqnarray}\label{resetting2}
\frac{\partial}{\partial t}\bar{\rho}(x,t)=D\frac{\partial}{\partial x}\left\{ |x|^{-\eta}\frac{\partial}{\partial x} \bar{\rho}(x,t)\right\}+re^{tr}\delta(x-x').
\end{eqnarray}
We then solve this new equation for $\bar{\rho}(x,t)$ using the Green function approach. In particular, the Green function ${\cal{G}}(x,\bar{x},t)$ related to Eq.~(\ref{resetting2}) is obtained by solving
\begin{eqnarray}
\label{Green_Resetting_1}
\frac{\partial}{\partial t}{\cal{G}}(x,\bar{x},t)-D\frac{\partial}{\partial x}\left\{ |x|^{-\eta}\frac{\partial}{\partial x} {\cal{G}}(x,\bar{x},t)\right\}=\delta(x-\bar{x})\delta(t),
\end{eqnarray}
subject to the conditions ${\cal{G}}(x,x',t)=0$ for $t<0$, and ${\cal{G}}(\pm \infty,x',t)=0$. In this approach, the distribution $\bar{\rho}(x,t)$ is given by
\begin{eqnarray}
\label{sol1}
\bar{\rho}(x,t)=\int_{-\infty}^{\infty}d\bar{x}\varphi(\bar{x}){\cal{G}}(x,\bar{x},t)+r\int_{0}^{t}dt'e^{r t'}{\cal{G}}(x,\bar{x},t-t'),
\end{eqnarray}
which in turn formally determines $\rho(x,t)$. 

We find the Green function of Eq.~(\ref{Green_Resetting_1}) by using the eigenfunctions of the Sturm--Liouville problem related to the following differential equation:
\begin{equation}
\label{SL1}
\frac{\partial}{\partial x}\left\{|x|^{-\eta}\frac{\partial}{\partial x}\psi\left(x,k\right)\right\}=-k^{2}\psi\left(x,k\right).
\end{equation}
Subjected to the boundary condition $\psi(\pm\infty,k)=0$, these eigenfunctions are
\begin{eqnarray}
\label{eigenfunction1}
\psi_{+}(x,k)=|x|^{\frac{1}{2}(1+\eta)}{\mbox{J}}_{-\nu}\left(\frac{2k|x|^{\frac{1}{2}(2+\eta)}}{2+\eta}\right)
\end{eqnarray}
and
\begin{eqnarray} \label{eigenfunction2}\psi_{-}(x,k)=x|x|^{\frac{1}{2}(1+\eta)-1}{\mbox{J}}_{\nu}\left(\frac{2k|x|^{\frac{1}{2}(2+\eta)}}{2+\eta}\right),
\end{eqnarray}
where $\nu=(1+\eta)/(2+\eta)$ and ${\mbox{J}}_{\nu}(x)$ is the Bessel function~\cite{book9895}. Using the eigenfunctions defined by Eqs.~(\ref{eigenfunction1}) and~(\ref{eigenfunction2}), the Green function can be written as
\begin{eqnarray}
\label{Green2}
{\cal{G}}(x,\bar{x},t)=\left.\left.\frac{2}{2+\eta}\int_{0}^{\infty}dkk\right[\psi_{+}(x,k)\tilde{{\cal{G}}}_{+}(k,\bar{x},t)+\psi_{-}(x,k)\tilde{{\cal{G}}}_{-}(k,\bar{x},t)\right], 
\end{eqnarray}
with
\begin{eqnarray}
\tilde{{\cal{G}}}_{\pm}(k,\bar{x},t) = \frac{1}{2}\int_{-\infty}^{\infty}dx \psi_{\pm}(x,k){\cal{G}}(x,\bar{x},t)\;,
\end{eqnarray}
where ${\cal{G}}_{\pm}(k,\bar{x},t)$ is determined by Eq.~(\ref{Green_Resetting_1}). 

\begin{figure}[ht]
\centering
\includegraphics[width=1\columnwidth]{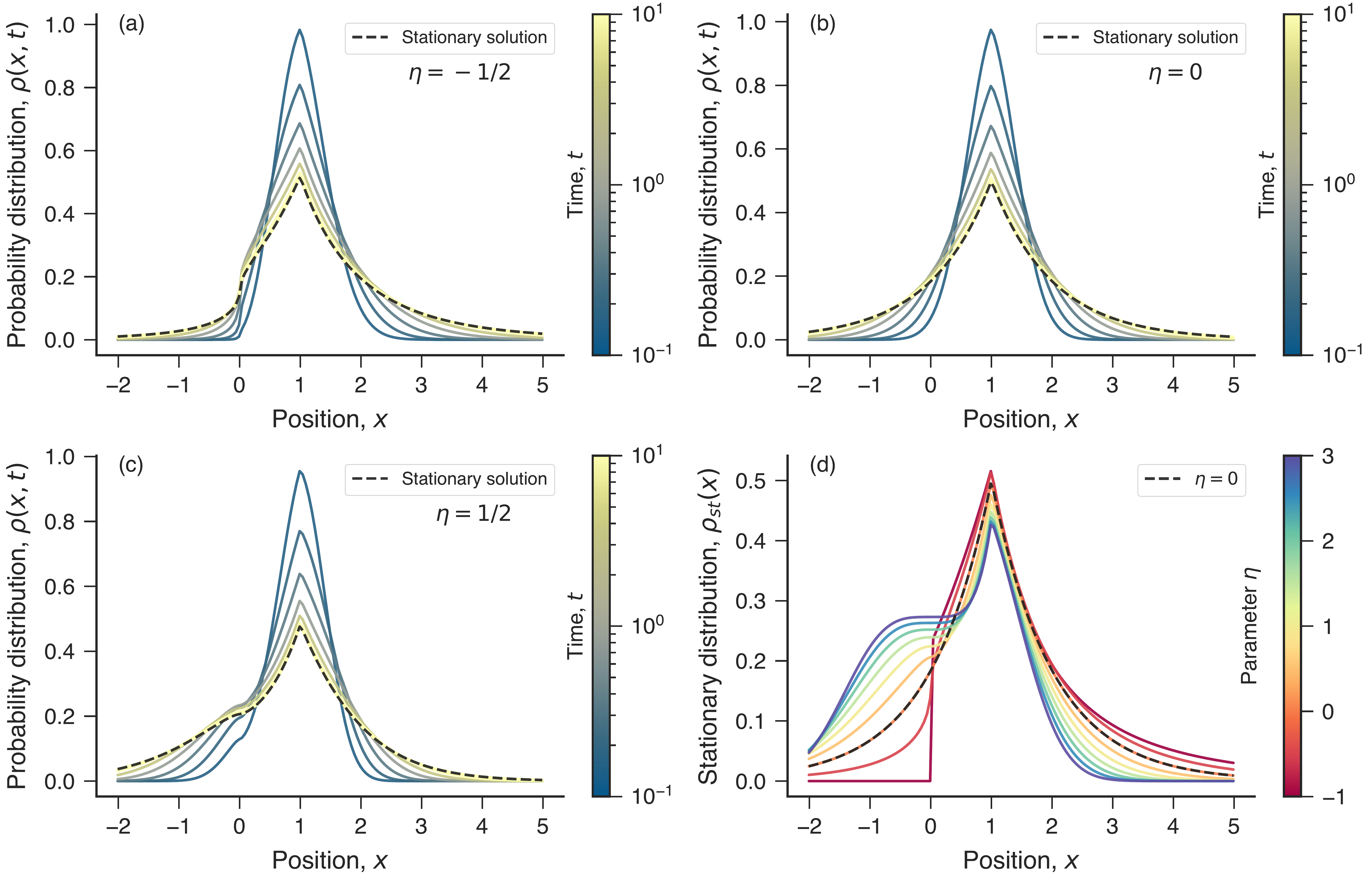}
\caption{Evolution of the probability distributions and the stationary states of a diffusion process on heterogeneous media with stochastic resetting. Panels (a), (b), and (c) show the probability distribution defined by Eq.~(\ref{sol2}) for different values of $\eta$ (indicated within each panel) and $t$ (indicated by the color bar). In these panels, the dashed lines represent the stationary distributions. Panel (c) shows the stationary distribution defined by Eqs.~(\ref{sol2_est1}) and ~(\ref{sol2_est2}) for different values of $\eta$ (indicated by the color bar). We have considered $x'=1$, $D=1$, and $\varphi(x)=\delta(x-x')$ in all panels.}
\label{Figure1}
\end{figure}

By substituting Eq.~(\ref{Green2}) into Eq.~(\ref{Green_Resetting_1}) and using the orthogonality of the eigenfunctions, we find that  
\begin{eqnarray}
\label{Green_Resetting_2}
\frac{\partial}{\partial t}\tilde{{\cal{G}}}_{\pm}(k,\bar{x},t)+
k^{2}D\tilde{{\cal{G}}}_{\pm}(k,\bar{x},t) =\frac{1}{2}\psi_{\pm}(\bar{x},k)\delta(t)\;,
\end{eqnarray}
whose solution is
\begin{eqnarray}
\label{Green_Resetting_3}
\tilde{{\cal{G}}}_{\pm}(k,\bar{x},t)&=&\frac{1}{2}\psi_{\pm}(\bar{x},k)e^{-Dk^{2}t}\;.
\end{eqnarray}
For $t>0$, the substitution Eq.~(\ref{Green_Resetting_3}) into Eq.~(\ref{Green2}) yields
\begin{eqnarray}
{\cal{G}}(x,\bar{x},t)=\left.\left.\frac{1}{2+\eta}\int_{0}^{\infty}dkk\right[\!\!\psi_{+}(x,k)\psi_{+}(\bar{x},k)+\psi_{-}(x,k)\psi_{-}(\bar{x},k)\!\right]e^{-Dk^{2}t} \;,
\end{eqnarray}
and after some calculations, we finally find
\begin{eqnarray}
\label{Green_Resetting_final}
{\cal{G}}(x,\bar{x},t)&=&\frac{|x\bar{x}|^{\frac{1}{2}\left(1+\eta\right)}}{2(2+\eta)Dt}e^{-\frac{|x|^{2+\eta}+|\bar{x}|^{2+\eta}}{(2+\eta)^{2}Dt}}\nonumber \\ &\times&\left\{{\mbox{I}}_{- \nu}\left(\frac{2|x\bar{x}|^{\frac{1}{2}\left(2+\eta\right)}}
{(2+\eta)^{2}Dt}\right)+\frac{x\bar{x}}{|x\bar{x}|}{\mbox{I}}_{\nu}\left(\frac{2|x\bar{x}|^{\frac{1}{2}\left(2+\eta\right)}}
{(2+\eta)^{2}Dt}\right)\right\}\;,
\end{eqnarray}
where ${\mbox{I}}_{\nu}(x)$ is the Bessel function of modified argument of first kind~\cite{book9895}.

Now that we have found the Green function [Eq.~(\ref{Green_Resetting_final})], the probability distribution associated with our diffusion equation [Eq.~(\ref{resetting1})] can be written as
\begin{eqnarray}
\label{sol2}
\rho(x,t)=e^{-rt}\int_{-\infty}^{\infty}d\bar{x}\varphi(\bar{x}){\cal{G}}(x,\bar{x},t)+ r\int_{0}^{t}dt'e^{-rt'}{\cal{G}}(x,x',t').
\end{eqnarray}
Figures~\ref{Figure1}(a), \ref{Figure1}(b) and \ref{Figure1}(c) illustrate the time dependent behavior of the distribution $\rho(x,t)$ for three values of $\eta$. In these figures, we have numerically integrated Eq.~(\ref{sol2}) by considering $x'=1$, $D=1$, and $\varphi(x)=\delta(x-x')$. We note that shape of these distributions dependent on the parameter $\eta$ associated with the media heterogeneity and that they significant differ from the homogeneous case ($\eta=0$). Furthermore, the diminishing differences between these distributions for increasing values of $t$ indicate the they are approaching a stationary state. Indeed, by considering the limit $t\rightarrow\infty$ and integrating Eq.~(\ref{sol2}), we find that this stationary distribution is given by
\begin{eqnarray}\label{sol2_est1}
\rho_{st}(x)&=&\frac{r|xx'|^{\frac{1}{2}(1+\eta)}}{(2+\eta)D}\left[{\mbox{I}}_{-\nu}\left(\frac{2}{2+\eta}\sqrt{\frac{r}{D}}|x|^{\frac{1}{2}(2+\eta)}\right){\mbox{K}}_{-\nu}\left(\frac{2}{2+\eta}\sqrt{\frac{r}{D}}|x'|^{\frac{1}{2}(2+\eta)}\right)\right.\nonumber \\ &+& \left. \frac{xx'}{|xx'|}{\mbox{I}}_{\nu}\left(\frac{2}{2+\eta}\sqrt{\frac{r}{D}}|x|^{\frac{1}{2}(2+\eta)}\right){\mbox{K}}_{\nu}\left(\frac{2}{2+\eta}\sqrt{\frac{r}{D}}|x'|^{\frac{1}{2}(2+\eta)}\right)\right]\;,
\end{eqnarray}
for $|x|<|x'|$, and 
\begin{eqnarray}\label{sol2_est2}
\rho_{st}(x)&=&\frac{r|xx'|^{\frac{1}{2}(1+\eta)}}{(2+\eta)D}\left[{\mbox{I}}_{-\nu}\left(\frac{2}{2+\eta}\sqrt{\frac{r}{D}}|x'|^{\frac{1}{2}(2+\eta)}\right){\mbox{K}}_{-\nu}\left(\frac{2}{2+\eta}\sqrt{\frac{r}{D}}|x|^{\frac{1}{2}(2+\eta)}\right) \right. \nonumber\\ 
&+&\left. \frac{xx'}{|xx'|}{\mbox{I}}_{\nu}\left(\frac{2}{2+\eta}\sqrt{\frac{r}{D}}|x'|^{\frac{1}{2}(2+\eta)}\right){\mbox{K}}_{\nu}\left(\frac{2}{2+\eta}\sqrt{\frac{r}{D}}|x|^{\frac{1}{2}(2+\eta)}\right)\right]\;,
\end{eqnarray}
for $|x|>|x'|$, where ${\mbox{K}}_{\nu}(x)$ is the Bessel function of modified argument of second kind~\cite{book9895}. The dashed lines shown in Figs.~\ref{Figure1}(a), \ref{Figure1}(b) and \ref{Figure1}(c) represent this stationary distribution and Fig.~\ref{Figure1}(d) compares the shape of these stationary distributions for different values of $\eta$. We observe that the stationary states are strongly affected by the parameter $\eta$ and thus by media heterogeneity.

\begin{figure}[!t]
\centering
\includegraphics[width=1\columnwidth]{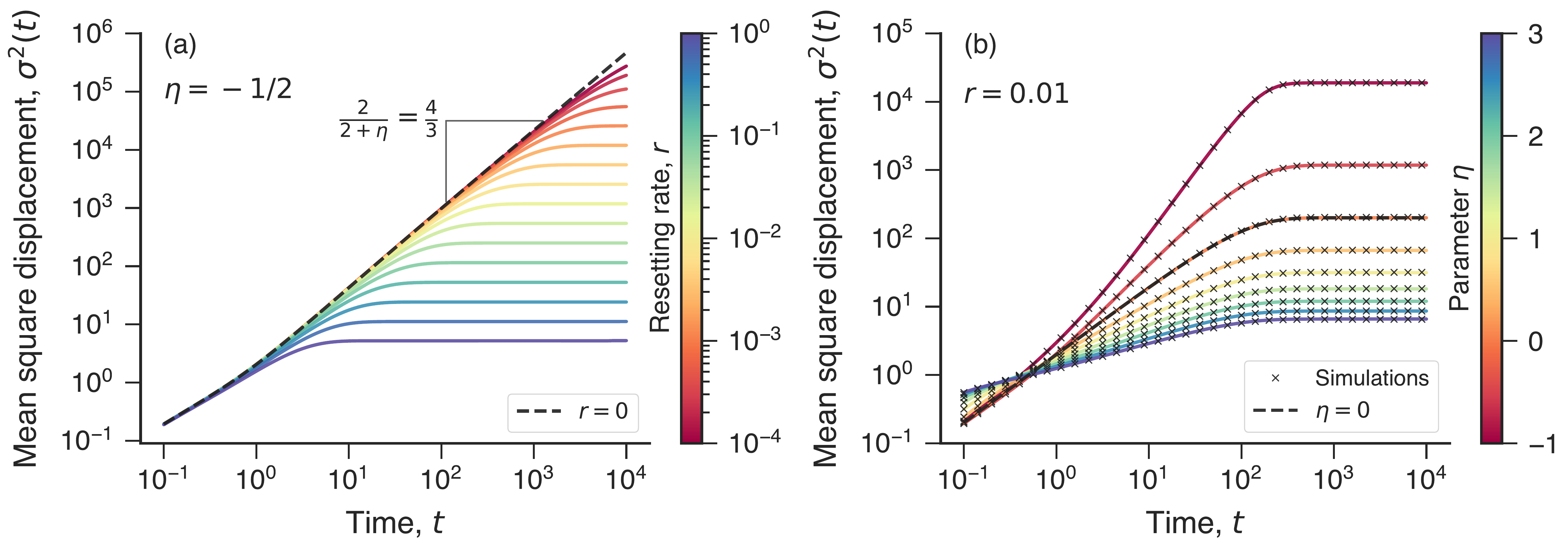}
\caption{Transient anomalous diffusion and the non-trivial dependence on the resetting rate and the media heterogeneity. (a) Time dependence of the mean square displacement $\sigma^2(t)$ for $\eta=-1/2$ and different resetting rates (indicated by the color bar). The dashed line indicates the diffusive behavior without stochastic resetting ($r=0$). (b) Time dependent behavior of the mean square displacement $\sigma^2(t)$ for $r=0.01$ and different values of the parameter $\eta$ (indicated by the color bar). The dashed line represent the homogeneous media case ($\eta=0$) and the markers indicate the results obtained via numerical simulations of a Langevin equation. In both panels, we have used $x'=1$, $D=1$, and $\varphi(x)=\delta(x-x')$.}
\label{Figure2}
\end{figure}

In addition to the probability distributions, we can also numerically calculate the mean square displacement $\sigma^2(t)=\langle\left(x-\langle x \rangle\right)^2\rangle$ by using Eq.~(\ref{sol2}). Without stochastic resetting ($r=0$), the mean square displacement is given by $\sigma^{2}(t)\propto t^{\frac{2}{2+\eta}}$~\cite{Book2018} and so we have sub-diffusion for $\eta>0$ and superdiffusion for $\eta<0$. As we have verified, stochastic resetting yields stationary states such that these anomalous regimes become a transient behavior before the system approaches the steady-state. To verify this behavior, we calculate the $\sigma^2(t)$ for different values of the resetting rate $r$, $\eta=-1/2$, $x'=1$, $D=1$, and $\varphi(x)=\delta(x-x')$. Figure~\ref{Figure2}(a) shows these results where the dashed line indicate the case without stochastic resetting ($r=0$). We observe that the mean square displacement rapidly saturates for high resetting rates. However, with the decrease of the resetting rate, the mean square displacement displays a transient anomalous diffusive regime characterized by $\sigma^{2}(t)\propto t^{\frac{2}{2+\eta}}$ before approaching the stead-state. The saturation plateau of the mean square displacement and the length of the transient regime increase with the decrease of the resetting rate. We have further calculated the mean square displacement for different values of $\eta$ while keeping the resetting rate constant ($r=0.01$). Figure~\ref{Figure2}(b) shows that the saturation plateau also depends on the parameter $\eta$ such that the larger the value of $\eta$, the lower the plateau. It is worth noticing that the mean square square displacement further exhibits an initial transient behavior for very short time scales [$t\lesssim1$ in Figs.~\ref{Figure2}(a) and \ref{Figure2}(b)]. This initial transient behavior also depends on $r$ and $\eta$, and we have verified that it only emerges for $x'\neq 0$, that is, we only observe the regime $\sigma^{2}(t)\propto t^{\frac{2}{2+\eta}}$ before the stead-state if $x'=0$.

To further strengthen our results, we have considered a Langevin equation~\cite{book12395} with suitable coefficients to numerically simulate the exact results obtained by solving Eq.~(\ref{resetting1}) with the Green function approach. We have focused on the behavior of the mean square displacement and adapted the approach presented in Ref.~\cite{evans2014diffusion} to account for the position-dependent diffusion coefficient [$D(x) = D|x|^{-\eta}$]. This Langevin equation describes a diffusive motion subjected renewal process, in which particle's position is reset to position $x'$ at a Poisson rate $r$. We have numerically simulated this Langevin equation for $x'=1$,  $D=1$, $r=0.01$, initial position $x(t=0)=1$, and different values of $\eta\in\{-1, -1/2, 0, \dots, 3\}$ up to the maximum integration time $t_{\text{max}}=10^4$. Using these simulations, we create an ensemble with one thousand trajectories for each value of $\eta$, from which we estimate the mean square displacement. Figure~\ref{Figure2}(b) shows these values (black markers) where we observe an excellent agreement between the simulations and the exact form of the mean square displacement calculated using Eq.~(\ref{sol2}).

\section{first-passage time on heterogeneous media with stochastic resetting}

In addition to the probability distributions and the mean square displacement, we have also investigated the first-passage time problem associated with the diffusive process described by Eq.~(\ref{resetting1}). To do so, we use the backward master equation approach for the survival probability ${\cal{Q}}(x,t)$ and assume the initial position $x>0$ to be a variable and the resetting position $x'=x_0$, that is, 
\begin{eqnarray}
\label{master}
\frac{\partial }{\partial t}{\cal{Q}}(x,t)=\frac{\partial}{\partial x}\left\{D|x|^{-\eta}\frac{\partial}{\partial x}{\cal{Q}}(x,t)\right\}-r {\cal{Q}}(x,t)+r {\cal{Q}}(x_{0}, t)
\end{eqnarray}
with boundary conditions ${\cal{Q}}(0, t)=0$ and ${\cal{Q}}(\infty,t)=0$, and initial condition ${\cal{Q}}(x, 0)=1$.  This equation represents the probability that a particle starting at the position $x$ does not reach the origin ($x=0$) until the time $t$. Furthermore, the second and third terms on the right side of this equation correspond to the resetting of the initial position from $x$ to $x_{0}$, which implies a loss of probability from ${\cal{Q}}(x, t)$ and a gain of probability to ${\cal{Q}}\left(x_{0}, t\right)$.

To solve Eq.~(\ref{master}), we first apply Laplace transform ($\widetilde{{\cal{Q}}}(x, s)=$ ${\cal{L}}\left\{ {\cal{Q}}(x, t);s\right\}$), yielding 
\begin{eqnarray}
\frac{\partial}{\partial x}\left\{D|x|^{-\eta}\frac{\partial}{\partial x}\widetilde{{\cal{Q}}}(x,s)\right\}-(r+s) \widetilde{{\cal{Q}}}(x,s)=-1-r \widetilde{{\cal{Q}}}(x_{0},s)\,,
\end{eqnarray}
which has the general solution
\begin{eqnarray}
\widetilde{{\cal{Q}}}(x, s)&=&{\cal{A}}x^{\frac{1}{2}(1+\eta)} {\mathrm{I}}_{\nu}\left(\frac{2\alpha}{2+\eta}x^{\frac{1}{2}(2+\eta)}\right)+ {\cal{B}}x^{\frac{1}{2}(1+\eta)}{\mathrm{K}}_{\nu}\left(\frac{2\alpha}{2+\eta}x^{\frac{1}{2}(2+\eta)}\right)\nonumber \\ &+&\left(1+r {\cal{Q}}\left(x_{0}, s\right)\right) /(r+s), 
\end{eqnarray}
where $\cal{A}$ and $\cal{B}$ are integrating constants and $\alpha=\sqrt{(r+s) / D}$. The boundary condition ${\cal{Q}}(\infty,s)=0$ yields ${\cal{A}}=0$, while $\widetilde{{\cal{Q}}}(0, s)=0$ implies
\begin{eqnarray}
{\cal{B}}=-\frac{1+r\widetilde{{\cal{Q}}}\left(x_{0}, s\right)}{s+r}\left[1-\frac{2}{\Gamma\left(\frac{1+\eta}{2+\eta}\right)}\left(\frac{\alpha}{2+\eta}\right)^{\frac{1+\eta}{2+\eta}}x_{0}^{\frac{1}{2}(1+\eta)}{\mathrm{K}}_{\nu}\left(\frac{2\alpha}{2+\eta}x_{0}^{\frac{1}{2}(2+\eta)}\right)\right]\,,
\end{eqnarray}
where $\widetilde{{\cal{Q}}}\left(x_{0}, s\right)$ is determined self-consistently, leading to
\begin{eqnarray}
\widetilde{{\cal{Q}}}(x_{0}, s)=\frac{1-\frac{2}{\Gamma\left(\frac{1+\eta}{2+\eta}\right)}\left(\frac{\alpha}{2+\eta}\right)^{\frac{1+\eta}{2+\eta}}x_{0}^{\frac{1}{2}(1+\eta)}{\mathrm{K}}_{\nu}\left(\frac{2\alpha}{2+\eta}x_{0}^{\frac{1}{2}(2+\eta)}\right)}{s+ \frac{2r}{\Gamma\left(\frac{1+\eta}{2+\eta}\right)}\left(\frac{\alpha}{2+\eta}\right)^{\frac{1+\eta}{2+\eta}}x_{0}^{\frac{1}{2}(1+\eta)}{\mathrm{K}}_{\nu}\left(\frac{2\alpha}{2+\eta}x_{0}^{\frac{1}{2}(2+\eta)}\right)}.
\end{eqnarray}

By considering the initial position to be equal to the resetting position ($x=x_0$), the mean first-passage time $T\left(x_{0}\right)$ can be calculated from
\begin{eqnarray}
T\left(x_{0}\right)&=&-\int_{0}^{\infty} dt\; t \frac{\partial}{\partial t}  {\cal{Q}}\left(x_{0}, t\right) =\widetilde{{\cal{Q}}}\left(x_{0}, 0\right)\,,
\end{eqnarray}
and yields
\begin{eqnarray}\label{avgfpt}
T\left(x_{0}\right)&=&\frac{1-\frac{2}{\Gamma\left(\frac{1+\eta}{2+\eta}\right)}\left(\frac{\bar{\alpha}}{2+\eta}\right)^{\frac{1+\eta}{2+\eta}}x_{0}^{\frac{1}{2}(1+\eta)}{\mathrm{K}}_{\nu}\left(\frac{2\bar{\alpha}}{2+\eta}x_{0}^{\frac{1}{2}(2+\eta)}\right)}{ \frac{2r}{\Gamma\left(\frac{1+\eta}{2+\eta}\right)}\left(\frac{\bar{\alpha}}{2+\eta}\right)^{\frac{1+\eta}{2+\eta}}x_{0}^{\frac{1}{2}(1+\eta)}{\mathrm{K}}_{\nu}\left(\frac{2\bar{\alpha}}{2+\eta}x_{0}^{\frac{1}{2}(2+\eta)}\right)}\;,
\end{eqnarray}
where $\bar{\alpha}=\sqrt{r /D}$. This quantity thus represents the average time to reach the position $x=0$ (origin) spent by a particle starting at $x=x_0$. We notice that $T(x_{0})$ is finite for $r>0$ and diverges when $r \rightarrow 0$ (that is, without stochastic resetting). Indeed, we have $T \sim 1/r^{\frac{1}{2+\eta}}$ in the limit $r \rightarrow 0$, recovering the well-known behavior of the mean first-passage time associated with Brownian motion for $\eta=0$. It is also worth noticing that $T$ diverges as $r \rightarrow \infty$; that is, if the resetting rate increases too much, the diffusing particle has less time between resets to reach the origin. We further recover the result of Ref.~\cite{evans2011diffusion} obtained for homogeneous media when $\eta=0$. 

\begin{figure}[!t]
\centering
\includegraphics[width=1\columnwidth]{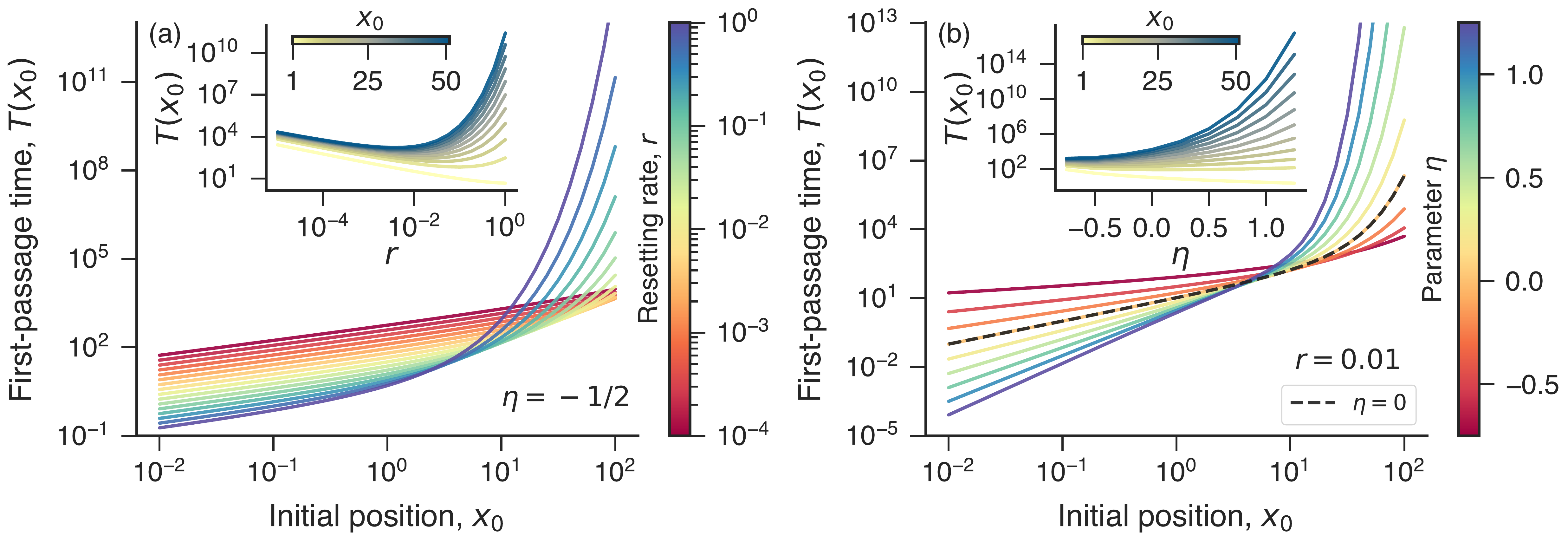}
\caption{Dependence of the mean first-passage time on the resetting rate and the media heterogeneity. (a) Mean first-passage time $T(x_0)$ as a function of the initial position $x_0$ [Eq.~(\ref{avgfpt})] for $\eta=-1/2$, $D=1$ and different values of the resetting rate $r$ (indicated by the color bar). The inset represents the values of $T(x_0)$ as a function of $r$ for different values of $x_0$ (indicated by the color bar). (a) Mean first-passage time $T(x_0)$ as a function of the initial position $x_0$ [Eq.~(\ref{avgfpt})] for $r=0.01$ and different values of the parameter $\eta$. In this panel, the dashed line represents the case of homogeneous media ($\eta=0$). The inset shows the values of $T(x_0)$ as a function of $\eta$ for different values of $x_0$ (indicated by the color bar).}
\label{Figure3}
\end{figure}

Figure~\ref{Figure3}(a) shows the mean first-passage time $T(x_0)$ as a function of the initial position $x_0$ for $\eta=-1/2$ and different values of the resetting rate $r$. When the initial position is very close to the origin, we observe that an increase in the values $r$ associates with a reduction in $T(x_0)$. Conversely, a rise in the resetting rate $r$ implies an increase in $T(x_0)$ when the particle's initial position is far from the origin. This behavior implies the existence of an optimal resetting rate (that also depends on $\eta$) for which the mean first-passage time $T(x_0)$ is minimum [see inset of Fig.~\ref{Figure3}(a)]. In addition to depending on the resetting rate $r$, we observe that the mean first-passage time given by Eq.~(\ref{avgfpt}) is affected by the parameter $\eta$ and so by the media heterogeneity. Figure~\ref{Figure3}(b) depicts the values of $T(x_0)$ as a function of $x_0$ with resetting rate $r=0.01$ and different values of the parameter $\eta$. In this figure, the dashed line indicates the results for homogeneous media ($\eta=0$). For initial positions close to the origin, an increase in $\eta$ is followed by a decrease in the mean first-passage time. However, as the initial position moves away from the origin, the effect of $\eta$ on $T(x_0)$ is reversed; that is, the mean first-passage time increases with the rise of $\eta$ [see inset of Fig.~\ref{Figure3}(b)].

\section{Discussion and Conclusions}

We have investigated a diffusive process on heterogeneous media with stochastic resetting. The system heterogeneity is modelled by considering a spatial-dependent diffusion coefficient that depends on particles' positions via a power-law function -- a dependence that has been used in several problems related to anomalous diffusion~\cite{PhysRevLett.54.455,PhysRevA.32.3073,PhysRevE.60.5528,richardson1926atmospheric,book1371194,book1371197}. We have obtained exact solutions for the probability distribution of particles' positions using the Green function approach which in turn allowed us to calculate the temporal dependence of the mean square displacement. We have also estimated the mean square displacement using simulations of a Langevin equation, finding an excellent agreement between numerical and exact results. Results demonstrate that our system exhibit a transient anomalous diffusion (with superdiffusive or sub-diffusive regimes) before reaching a stationary state that depends both on the resetting rate ($r$) and the parameter controlling the media heterogeneity ($\eta$). This result contrast with case without heterogeneity ($\eta=0$) for which the system presents usual diffusion before reaching a stead-state. 

We have also studied the first-passage time problem related to this diffusive process and obtained an exact expression for the mean first-passage time. Similarly to what happens for the mean square displacement, we observe that the mean first-passage time non-trivially depends on the resetting rate ($r$) and the parameter controlling the media heterogeneity ($\eta$). Intriguingly, we have found an optimal resetting rate for which the mean first-passage time presents a minimum that is further dependent on the media heterogeneity ($\eta$). We have also found that the effect of the parameter $\eta$ on the mean first-passage time changes depending on the initial distance to the target. The mean first-passage time decreases with $\eta$ for small distances, while this quantity increase with $\eta$ for large distances. When taking together, our results for the mean square displacement and the mean first-passage time indicate that the media heterogeneity significantly affect the diffusive properties a system with stochastic resetting, yielding a rich and flexible framework that may eventually describes real-world processes. 

\acknowledgments
This work was partially supported by the CNPq (Brazilian Agency). E.K.L. acknowledges the support of the Conselho Nacional de Desenvolvimento Cient\'ifico e Tecnol\'ogico (CNPq -- Grant 302983/2018-0 ). H.V.R. acknowledges the support of the Conselho Nacional de Desenvolvimento Cient\'ifico e Tecnol\'ogico (CNPq -- Grants 407690/2018-2 and 303121/2018-1).

\bibliography{refer.bib}

%apsrev4-2.bst 2019-01-14 (MD) hand-edited version of apsrev4-1.bst
%Control: key (0)
%Control: author (8) initials jnrlst
%Control: editor formatted (1) identically to author
%Control: production of article title (0) allowed
%Control: page (0) single
%Control: year (1) truncated
%Control: production of eprint (0) enabled
\begin{thebibliography}{45}%
\makeatletter
\providecommand \@ifxundefined [1]{%
 \@ifx{#1\undefined}
}%
\providecommand \@ifnum [1]{%
 \ifnum #1\expandafter \@firstoftwo
 \else \expandafter \@secondoftwo
 \fi
}%
\providecommand \@ifx [1]{%
 \ifx #1\expandafter \@firstoftwo
 \else \expandafter \@secondoftwo
 \fi
}%
\providecommand \natexlab [1]{#1}%
\providecommand \enquote  [1]{``#1''}%
\providecommand \bibnamefont  [1]{#1}%
\providecommand \bibfnamefont [1]{#1}%
\providecommand \citenamefont [1]{#1}%
\providecommand \href@noop [0]{\@secondoftwo}%
\providecommand \href [0]{\begingroup \@sanitize@url \@href}%
\providecommand \@href[1]{\@@startlink{#1}\@@href}%
\providecommand \@@href[1]{\endgroup#1\@@endlink}%
\providecommand \@sanitize@url [0]{\catcode `\\12\catcode `\$12\catcode
  `\&12\catcode `\#12\catcode `\^12\catcode `\_12\catcode `\%12\relax}%
\providecommand \@@startlink[1]{}%
\providecommand \@@endlink[0]{}%
\providecommand \url  [0]{\begingroup\@sanitize@url \@url }%
\providecommand \@url [1]{\endgroup\@href {#1}{\urlprefix }}%
\providecommand \urlprefix  [0]{URL }%
\providecommand \Eprint [0]{\href }%
\providecommand \doibase [0]{https://doi.org/}%
\providecommand \selectlanguage [0]{\@gobble}%
\providecommand \bibinfo  [0]{\@secondoftwo}%
\providecommand \bibfield  [0]{\@secondoftwo}%
\providecommand \translation [1]{[#1]}%
\providecommand \BibitemOpen [0]{}%
\providecommand \bibitemStop [0]{}%
\providecommand \bibitemNoStop [0]{.\EOS\space}%
\providecommand \EOS [0]{\spacefactor3000\relax}%
\providecommand \BibitemShut  [1]{\csname bibitem#1\endcsname}%
\let\auto@bib@innerbib\@empty
%</preamble>
\bibitem [{\citenamefont {Einstein}(1905)}]{einstein}%
  \BibitemOpen
  \bibfield  {author} {\bibinfo {author} {\bibfnamefont {A.}~\bibnamefont
  {Einstein}},\ }\bibfield  {title} {\bibinfo {title} {{\"U}ber die von der
  molekularkinetischen theorie der w{\"a}rme geforderte bewegung von in
  ruhenden fl{\"u}ssigkeiten suspendierten teilchen},\ }\href@noop {}
  {\bibfield  {journal} {\bibinfo  {journal} {Ann. Phys.}\ }\textbf {\bibinfo
  {volume} {322}},\ \bibinfo {pages} {549} (\bibinfo {year}
  {1905})}\BibitemShut {NoStop}%
\bibitem [{\citenamefont {Langevin}(1908)}]{langevin}%
  \BibitemOpen
  \bibfield  {author} {\bibinfo {author} {\bibfnamefont {P.}~\bibnamefont
  {Langevin}},\ }\bibfield  {title} {\bibinfo {title} {Sur la th{\'e}orie du
  mouvement brownien},\ }\href@noop {} {\bibfield  {journal} {\bibinfo
  {journal} {CR Acad. Sci. Paris}\ }\textbf {\bibinfo {volume} {146}},\
  \bibinfo {pages} {530} (\bibinfo {year} {1908})}\BibitemShut {NoStop}%
\bibitem [{\citenamefont {Smoluchowski}(1906)}]{von1906}%
  \BibitemOpen
  \bibfield  {author} {\bibinfo {author} {\bibfnamefont {M.~V.}\ \bibnamefont
  {Smoluchowski}},\ }\bibfield  {title} {\bibinfo {title} {Zur kinetischen
  theorie der brownschen molekularbewegung und der suspensionen},\ }\href@noop
  {} {\bibfield  {journal} {\bibinfo  {journal} {Ann. Phys.}\ }\textbf
  {\bibinfo {volume} {326}},\ \bibinfo {pages} {756} (\bibinfo {year}
  {1906})}\BibitemShut {NoStop}%
\bibitem [{\citenamefont {Pearson}(1905)}]{KP}%
  \BibitemOpen
  \bibfield  {author} {\bibinfo {author} {\bibfnamefont {K.}~\bibnamefont
  {Pearson}},\ }\bibfield  {title} {\bibinfo {title} {The problem of the random
  walk},\ }\href@noop {} {\bibfield  {journal} {\bibinfo  {journal} {Nature}\
  }\textbf {\bibinfo {volume} {72}},\ \bibinfo {pages} {294} (\bibinfo {year}
  {1905})}\BibitemShut {NoStop}%
\bibitem [{\citenamefont {Evangelista}\ and\ \citenamefont
  {Lenzi}(2018)}]{Book2018}%
  \BibitemOpen
  \bibfield  {author} {\bibinfo {author} {\bibfnamefont {L.~R.}\ \bibnamefont
  {Evangelista}}\ and\ \bibinfo {author} {\bibfnamefont {E.~K.}\ \bibnamefont
  {Lenzi}},\ }\href@noop {} {\emph {\bibinfo {title} {Fractional Diffusion
  Equations and Anomalous Diffusion}}}\ (\bibinfo  {publisher} {Cambridge
  University Press},\ \bibinfo {year} {2018})\BibitemShut {NoStop}%
\bibitem [{\citenamefont {Lenzi}\ \emph {et~al.}(2017)\citenamefont {Lenzi},
  \citenamefont {Zola}, \citenamefont {Ribeiro}, \citenamefont {Vieira},
  \citenamefont {Ciuchi}, \citenamefont {Mazzulla}, \citenamefont
  {Scaramuzza},\ and\ \citenamefont {Evangelista}}]{ionmotion}%
  \BibitemOpen
  \bibfield  {author} {\bibinfo {author} {\bibfnamefont {E.~K.}\ \bibnamefont
  {Lenzi}}, \bibinfo {author} {\bibfnamefont {R.~S.}\ \bibnamefont {Zola}},
  \bibinfo {author} {\bibfnamefont {H.~V.}\ \bibnamefont {Ribeiro}}, \bibinfo
  {author} {\bibfnamefont {D.~S.}\ \bibnamefont {Vieira}}, \bibinfo {author}
  {\bibfnamefont {F.}~\bibnamefont {Ciuchi}}, \bibinfo {author} {\bibfnamefont
  {A.}~\bibnamefont {Mazzulla}}, \bibinfo {author} {\bibfnamefont
  {N.}~\bibnamefont {Scaramuzza}},\ and\ \bibinfo {author} {\bibfnamefont
  {L.~R.}\ \bibnamefont {Evangelista}},\ }\bibfield  {title} {\bibinfo {title}
  {Ion motion in electrolytic cells: Anomalous diffusion evidences},\
  }\href@noop {} {\bibfield  {journal} {\bibinfo  {journal} {J. Phys. Chem. B}\
  }\textbf {\bibinfo {volume} {121}},\ \bibinfo {pages} {2882} (\bibinfo {year}
  {2017})}\BibitemShut {NoStop}%
\bibitem [{\citenamefont {Lippincott-Schwartz}\ \emph
  {et~al.}(2001)\citenamefont {Lippincott-Schwartz}, \citenamefont {Snapp},\
  and\ \citenamefont {Kenworthy}}]{lippincott2001studying}%
  \BibitemOpen
  \bibfield  {author} {\bibinfo {author} {\bibfnamefont {J.}~\bibnamefont
  {Lippincott-Schwartz}}, \bibinfo {author} {\bibfnamefont {E.}~\bibnamefont
  {Snapp}},\ and\ \bibinfo {author} {\bibfnamefont {A.}~\bibnamefont
  {Kenworthy}},\ }\bibfield  {title} {\bibinfo {title} {Studying protein
  dynamics in living cells},\ }\href@noop {} {\bibfield  {journal} {\bibinfo
  {journal} {Nat. Rev. Mol. Cell Biol.}\ }\textbf {\bibinfo {volume} {2}},\
  \bibinfo {pages} {444} (\bibinfo {year} {2001})}\BibitemShut {NoStop}%
\bibitem [{\citenamefont {Albinali}\ \emph {et~al.}(2016)\citenamefont
  {Albinali}, \citenamefont {Ozkan} \emph {et~al.}}]{albinali2016anomalous}%
  \BibitemOpen
  \bibfield  {author} {\bibinfo {author} {\bibfnamefont {A.}~\bibnamefont
  {Albinali}}, \bibinfo {author} {\bibfnamefont {E.}~\bibnamefont {Ozkan}},
  \emph {et~al.},\ }\bibfield  {title} {\bibinfo {title} {Anomalous diffusion
  approach and field application for fractured nano-porous reservoirs},\ }in\
  \href@noop {} {\emph {\bibinfo {booktitle} {SPE Annual Technical Conference
  and Exhibition}}}\ (\bibinfo {organization} {Society of Petroleum
  Engineers},\ \bibinfo {year} {2016})\BibitemShut {NoStop}%
\bibitem [{\citenamefont {Razminia}\ \emph {et~al.}(2020)\citenamefont
  {Razminia}, \citenamefont {Razminia},\ and\ \citenamefont
  {Shiryaev}}]{RAZMINIA2020105068}%
  \BibitemOpen
  \bibfield  {author} {\bibinfo {author} {\bibfnamefont {K.}~\bibnamefont
  {Razminia}}, \bibinfo {author} {\bibfnamefont {A.}~\bibnamefont {Razminia}},\
  and\ \bibinfo {author} {\bibfnamefont {V.~I.}\ \bibnamefont {Shiryaev}},\
  }\bibfield  {title} {\bibinfo {title} {Application of fractal geometry to
  describe reservoirs with complex structures},\ }\href@noop {} {\bibfield
  {journal} {\bibinfo  {journal} {Commun. Nonlinear Sci.}\ }\textbf {\bibinfo
  {volume} {82}},\ \bibinfo {pages} {105068} (\bibinfo {year}
  {2020})}\BibitemShut {NoStop}%
\bibitem [{\citenamefont {Metzler}\ \emph {et~al.}(1994)\citenamefont
  {Metzler}, \citenamefont {Gl{\"o}ckle},\ and\ \citenamefont
  {Nonnenmacher}}]{metzler1994fractional}%
  \BibitemOpen
  \bibfield  {author} {\bibinfo {author} {\bibfnamefont {R.}~\bibnamefont
  {Metzler}}, \bibinfo {author} {\bibfnamefont {W.~G.}\ \bibnamefont
  {Gl{\"o}ckle}},\ and\ \bibinfo {author} {\bibfnamefont {T.~F.}\ \bibnamefont
  {Nonnenmacher}},\ }\bibfield  {title} {\bibinfo {title} {Fractional model
  equation for anomalous diffusion},\ }\href@noop {} {\bibfield  {journal}
  {\bibinfo  {journal} {Physica A}\ }\textbf {\bibinfo {volume} {211}},\
  \bibinfo {pages} {13} (\bibinfo {year} {1994})}\BibitemShut {NoStop}%
\bibitem [{\citenamefont {Kinkhabwala}\ and\ \citenamefont
  {Bastiaens}(2010)}]{kinkhabwala2010spatial}%
  \BibitemOpen
  \bibfield  {author} {\bibinfo {author} {\bibfnamefont {A.}~\bibnamefont
  {Kinkhabwala}}\ and\ \bibinfo {author} {\bibfnamefont {P.~I.}\ \bibnamefont
  {Bastiaens}},\ }\bibfield  {title} {\bibinfo {title} {Spatial aspects of
  intracellular information processing},\ }\href@noop {} {\bibfield  {journal}
  {\bibinfo  {journal} {Curr. Opin. Genet. Dev.}\ }\textbf {\bibinfo {volume}
  {20}},\ \bibinfo {pages} {31} (\bibinfo {year} {2010})}\BibitemShut {NoStop}%
\bibitem [{\citenamefont {Capoulade}\ \emph {et~al.}(2011)\citenamefont
  {Capoulade}, \citenamefont {Wachsmuth}, \citenamefont {Hufnagel},\ and\
  \citenamefont {Knop}}]{capoulade2011quantitative}%
  \BibitemOpen
  \bibfield  {author} {\bibinfo {author} {\bibfnamefont {J.}~\bibnamefont
  {Capoulade}}, \bibinfo {author} {\bibfnamefont {M.}~\bibnamefont
  {Wachsmuth}}, \bibinfo {author} {\bibfnamefont {L.}~\bibnamefont
  {Hufnagel}},\ and\ \bibinfo {author} {\bibfnamefont {M.}~\bibnamefont
  {Knop}},\ }\bibfield  {title} {\bibinfo {title} {Quantitative fluorescence
  imaging of protein diffusion and interaction in living cells},\ }\href@noop
  {} {\bibfield  {journal} {\bibinfo  {journal} {Nat. Biotechnol.}\ }\textbf
  {\bibinfo {volume} {29}},\ \bibinfo {pages} {835} (\bibinfo {year}
  {2011})}\BibitemShut {NoStop}%
\bibitem [{\citenamefont {Weiss}\ \emph {et~al.}(0037)\citenamefont {Weiss},
  \citenamefont {Hashimoto},\ and\ \citenamefont {Nilsson}}]{Living1}%
  \BibitemOpen
  \bibfield  {author} {\bibinfo {author} {\bibfnamefont {M.}~\bibnamefont
  {Weiss}}, \bibinfo {author} {\bibfnamefont {H.}~\bibnamefont {Hashimoto}},\
  and\ \bibinfo {author} {\bibfnamefont {T.}~\bibnamefont {Nilsson}},\
  }\bibfield  {title} {\bibinfo {title} {Anomalous protein diffusion in living
  cells as seen by fluorescence correlation spectroscopy},\ }\href@noop {}
  {\bibfield  {journal} {\bibinfo  {journal} {Biophys. J.}\ }\textbf {\bibinfo
  {volume} {84}},\ \bibinfo {pages} {P4043} (\bibinfo {year}
  {20037})}\BibitemShut {NoStop}%
\bibitem [{\citenamefont {Gal}\ and\ \citenamefont {Weihs}(2010)}]{Living2}%
  \BibitemOpen
  \bibfield  {author} {\bibinfo {author} {\bibfnamefont {N.}~\bibnamefont
  {Gal}}\ and\ \bibinfo {author} {\bibfnamefont {D.}~\bibnamefont {Weihs}},\
  }\bibfield  {title} {\bibinfo {title} {Experimental evidence of strong
  anomalous diffusion in living cells},\ }\href@noop {} {\bibfield  {journal}
  {\bibinfo  {journal} {Phys. Rev. E}\ }\textbf {\bibinfo {volume} {81}},\
  \bibinfo {pages} {020903} (\bibinfo {year} {2010})}\BibitemShut {NoStop}%
\bibitem [{\citenamefont {Evans}\ and\ \citenamefont
  {Majumdar}(2011)}]{evans2011diffusion}%
  \BibitemOpen
  \bibfield  {author} {\bibinfo {author} {\bibfnamefont {M.~R.}\ \bibnamefont
  {Evans}}\ and\ \bibinfo {author} {\bibfnamefont {S.~N.}\ \bibnamefont
  {Majumdar}},\ }\bibfield  {title} {\bibinfo {title} {Diffusion with
  stochastic resetting},\ }\href@noop {} {\bibfield  {journal} {\bibinfo
  {journal} {Phys. Rev. Lett.}\ }\textbf {\bibinfo {volume} {106}},\ \bibinfo
  {pages} {160601} (\bibinfo {year} {2011})}\BibitemShut {NoStop}%
\bibitem [{\citenamefont {Evans}\ and\ \citenamefont
  {Majumdar}(2014)}]{evans2014diffusion}%
  \BibitemOpen
  \bibfield  {author} {\bibinfo {author} {\bibfnamefont {M.~R.}\ \bibnamefont
  {Evans}}\ and\ \bibinfo {author} {\bibfnamefont {S.~N.}\ \bibnamefont
  {Majumdar}},\ }\bibfield  {title} {\bibinfo {title} {Diffusion with resetting
  in arbitrary spatial dimension},\ }\href@noop {} {\bibfield  {journal}
  {\bibinfo  {journal} {J. Phys. A-Math. Theor.}\ }\textbf {\bibinfo {volume}
  {47}},\ \bibinfo {pages} {285001} (\bibinfo {year} {2014})}\BibitemShut
  {NoStop}%
\bibitem [{\citenamefont {Nagar}\ \emph {et~al.}(2011)\citenamefont {Nagar},
  \citenamefont {Valleriani},\ and\ \citenamefont
  {Lipowsky}}]{nagar2011translation}%
  \BibitemOpen
  \bibfield  {author} {\bibinfo {author} {\bibfnamefont {A.}~\bibnamefont
  {Nagar}}, \bibinfo {author} {\bibfnamefont {A.}~\bibnamefont {Valleriani}},\
  and\ \bibinfo {author} {\bibfnamefont {R.}~\bibnamefont {Lipowsky}},\
  }\bibfield  {title} {\bibinfo {title} {Translation by ribosomes with mrna
  degradation: Exclusion processes on aging tracks},\ }\href@noop {} {\bibfield
   {journal} {\bibinfo  {journal} {J. Stat. Phys.}\ }\textbf {\bibinfo {volume}
  {145}},\ \bibinfo {pages} {1385} (\bibinfo {year} {2011})}\BibitemShut
  {NoStop}%
\bibitem [{\citenamefont {Balaban}\ and\ \citenamefont
  {Luria}(2017)}]{balaban2017neural}%
  \BibitemOpen
  \bibfield  {author} {\bibinfo {author} {\bibfnamefont {H.}~\bibnamefont
  {Balaban}}\ and\ \bibinfo {author} {\bibfnamefont {R.}~\bibnamefont
  {Luria}},\ }\bibfield  {title} {\bibinfo {title} {Neural and behavioral
  evidence for an online resetting process in visual working memory},\
  }\href@noop {} {\bibfield  {journal} {\bibinfo  {journal} {J. Neurosci.}\
  }\textbf {\bibinfo {volume} {37}},\ \bibinfo {pages} {1225} (\bibinfo {year}
  {2017})}\BibitemShut {NoStop}%
\bibitem [{\citenamefont {Reuveni}\ \emph {et~al.}(2014)\citenamefont
  {Reuveni}, \citenamefont {Urbakh},\ and\ \citenamefont
  {Klafter}}]{Reuveni4391}%
  \BibitemOpen
  \bibfield  {author} {\bibinfo {author} {\bibfnamefont {S.}~\bibnamefont
  {Reuveni}}, \bibinfo {author} {\bibfnamefont {M.}~\bibnamefont {Urbakh}},\
  and\ \bibinfo {author} {\bibfnamefont {J.}~\bibnamefont {Klafter}},\
  }\bibfield  {title} {\bibinfo {title} {Role of substrate unbinding in
  michaelis{\textendash}menten enzymatic reactions},\ }\href
  {https://doi.org/10.1073/pnas.1318122111} {\bibfield  {journal} {\bibinfo
  {journal} {Proc. Natl. Acad. Sci.}\ }\textbf {\bibinfo {volume} {111}},\
  \bibinfo {pages} {4391} (\bibinfo {year} {2014})}\BibitemShut {NoStop}%
\bibitem [{\citenamefont {Bartumeus}\ and\ \citenamefont
  {Catalan}(2009)}]{bartumeus2009optimal}%
  \BibitemOpen
  \bibfield  {author} {\bibinfo {author} {\bibfnamefont {F.}~\bibnamefont
  {Bartumeus}}\ and\ \bibinfo {author} {\bibfnamefont {J.}~\bibnamefont
  {Catalan}},\ }\bibfield  {title} {\bibinfo {title} {Optimal search behavior
  and classic foraging theory},\ }\href@noop {} {\bibfield  {journal} {\bibinfo
   {journal} {J. Phys. A}\ }\textbf {\bibinfo {volume} {42}},\ \bibinfo {pages}
  {434002} (\bibinfo {year} {2009})}\BibitemShut {NoStop}%
\bibitem [{\citenamefont {Bhat}\ \emph {et~al.}(2016)\citenamefont {Bhat},
  \citenamefont {De~Bacco},\ and\ \citenamefont {Redner}}]{bhat2016stochastic}%
  \BibitemOpen
  \bibfield  {author} {\bibinfo {author} {\bibfnamefont {U.}~\bibnamefont
  {Bhat}}, \bibinfo {author} {\bibfnamefont {C.}~\bibnamefont {De~Bacco}},\
  and\ \bibinfo {author} {\bibfnamefont {S.}~\bibnamefont {Redner}},\
  }\bibfield  {title} {\bibinfo {title} {Stochastic search with poisson and
  deterministic resetting},\ }\href@noop {} {\bibfield  {journal} {\bibinfo
  {journal} {J. Stat. Mech.}\ }\textbf {\bibinfo {volume} {2016}},\ \bibinfo
  {pages} {083401} (\bibinfo {year} {2016})}\BibitemShut {NoStop}%
\bibitem [{\citenamefont {Evans}\ \emph {et~al.}(2013)\citenamefont {Evans},
  \citenamefont {Majumdar},\ and\ \citenamefont {Mallick}}]{Evans_2013}%
  \BibitemOpen
  \bibfield  {author} {\bibinfo {author} {\bibfnamefont {M.~R.}\ \bibnamefont
  {Evans}}, \bibinfo {author} {\bibfnamefont {S.~N.}\ \bibnamefont
  {Majumdar}},\ and\ \bibinfo {author} {\bibfnamefont {K.}~\bibnamefont
  {Mallick}},\ }\bibfield  {title} {\bibinfo {title} {Optimal diffusive search:
  nonequilibrium resetting versus equilibrium dynamics},\ }\href@noop {}
  {\bibfield  {journal} {\bibinfo  {journal} {J. Phys. A}\ }\textbf {\bibinfo
  {volume} {46}},\ \bibinfo {pages} {185001} (\bibinfo {year}
  {2013})}\BibitemShut {NoStop}%
\bibitem [{\citenamefont {Sandev}\ \emph {et~al.}(2021)\citenamefont {Sandev},
  \citenamefont {Domazetoski}, \citenamefont {Iomin},\ and\ \citenamefont
  {Kocarev}}]{sandev2021diffusion}%
  \BibitemOpen
  \bibfield  {author} {\bibinfo {author} {\bibfnamefont {T.}~\bibnamefont
  {Sandev}}, \bibinfo {author} {\bibfnamefont {V.}~\bibnamefont {Domazetoski}},
  \bibinfo {author} {\bibfnamefont {A.}~\bibnamefont {Iomin}},\ and\ \bibinfo
  {author} {\bibfnamefont {L.}~\bibnamefont {Kocarev}},\ }\bibfield  {title}
  {\bibinfo {title} {Diffusion--advection equations on a comb: Resetting and
  random search},\ }\href@noop {} {\bibfield  {journal} {\bibinfo  {journal}
  {Mathematics}\ }\textbf {\bibinfo {volume} {9}},\ \bibinfo {pages} {221}
  (\bibinfo {year} {2021})}\BibitemShut {NoStop}%
\bibitem [{\citenamefont {Pal}\ \emph {et~al.}(2016)\citenamefont {Pal},
  \citenamefont {Kundu},\ and\ \citenamefont {Evans}}]{pal2016diffusion}%
  \BibitemOpen
  \bibfield  {author} {\bibinfo {author} {\bibfnamefont {A.}~\bibnamefont
  {Pal}}, \bibinfo {author} {\bibfnamefont {A.}~\bibnamefont {Kundu}},\ and\
  \bibinfo {author} {\bibfnamefont {M.~R.}\ \bibnamefont {Evans}},\ }\bibfield
  {title} {\bibinfo {title} {Diffusion under time-dependent resetting},\
  }\href@noop {} {\bibfield  {journal} {\bibinfo  {journal} {J. Phys. A}\
  }\textbf {\bibinfo {volume} {49}},\ \bibinfo {pages} {225001} (\bibinfo
  {year} {2016})}\BibitemShut {NoStop}%
\bibitem [{\citenamefont {Pal}(2015)}]{pal2015diffusion}%
  \BibitemOpen
  \bibfield  {author} {\bibinfo {author} {\bibfnamefont {A.}~\bibnamefont
  {Pal}},\ }\bibfield  {title} {\bibinfo {title} {Diffusion in a potential
  landscape with stochastic resetting},\ }\href@noop {} {\bibfield  {journal}
  {\bibinfo  {journal} {Phys. Rev. E}\ }\textbf {\bibinfo {volume} {91}},\
  \bibinfo {pages} {012113} (\bibinfo {year} {2015})}\BibitemShut {NoStop}%
\bibitem [{\citenamefont {Evans}\ \emph {et~al.}(2020)\citenamefont {Evans},
  \citenamefont {Majumdar},\ and\ \citenamefont
  {Schehr}}]{evans2020stochastic}%
  \BibitemOpen
  \bibfield  {author} {\bibinfo {author} {\bibfnamefont {M.~R.}\ \bibnamefont
  {Evans}}, \bibinfo {author} {\bibfnamefont {S.~N.}\ \bibnamefont
  {Majumdar}},\ and\ \bibinfo {author} {\bibfnamefont {G.}~\bibnamefont
  {Schehr}},\ }\bibfield  {title} {\bibinfo {title} {Stochastic resetting and
  applications},\ }\href@noop {} {\bibfield  {journal} {\bibinfo  {journal} {J.
  Phys. A}\ }\textbf {\bibinfo {volume} {53}},\ \bibinfo {pages} {193001}
  (\bibinfo {year} {2020})}\BibitemShut {NoStop}%
\bibitem [{\citenamefont {Ray}\ and\ \citenamefont
  {Reuveni}(2020)}]{doi:10.1063/5.0010549}%
  \BibitemOpen
  \bibfield  {author} {\bibinfo {author} {\bibfnamefont {S.}~\bibnamefont
  {Ray}}\ and\ \bibinfo {author} {\bibfnamefont {S.}~\bibnamefont {Reuveni}},\
  }\bibfield  {title} {\bibinfo {title} {Diffusion with resetting in a
  logarithmic potential},\ }\href@noop {} {\bibfield  {journal} {\bibinfo
  {journal} {J. Chem. Phys.}\ }\textbf {\bibinfo {volume} {152}},\ \bibinfo
  {pages} {234110} (\bibinfo {year} {2020})}\BibitemShut {NoStop}%
\bibitem [{\citenamefont {Ray}\ and\ \citenamefont
  {Reuveni}(2021)}]{ray2021resetting}%
  \BibitemOpen
  \bibfield  {author} {\bibinfo {author} {\bibfnamefont {S.}~\bibnamefont
  {Ray}}\ and\ \bibinfo {author} {\bibfnamefont {S.}~\bibnamefont {Reuveni}},\
  }\bibfield  {title} {\bibinfo {title} {Resetting transition is governed by an
  interplay between thermal and potential energy},\ }\href@noop {} {\bibfield
  {journal} {\bibinfo  {journal} {J. Chem. Phys.}\ }\textbf {\bibinfo {volume}
  {154}},\ \bibinfo {pages} {171103} (\bibinfo {year} {2021})}\BibitemShut
  {NoStop}%
\bibitem [{\citenamefont {Shkilev}(2017)}]{shkilev2017continuous}%
  \BibitemOpen
  \bibfield  {author} {\bibinfo {author} {\bibfnamefont {V.}~\bibnamefont
  {Shkilev}},\ }\bibfield  {title} {\bibinfo {title} {Continuous-time random
  walk under time-dependent resetting},\ }\href@noop {} {\bibfield  {journal}
  {\bibinfo  {journal} {Phys. Rev. E}\ }\textbf {\bibinfo {volume} {96}},\
  \bibinfo {pages} {012126} (\bibinfo {year} {2017})}\BibitemShut {NoStop}%
\bibitem [{\citenamefont {dos Santos}(2019)}]{dos2019fractional}%
  \BibitemOpen
  \bibfield  {author} {\bibinfo {author} {\bibfnamefont {M.~A.}\ \bibnamefont
  {dos Santos}},\ }\bibfield  {title} {\bibinfo {title} {Fractional prabhakar
  derivative in diffusion equation with non-static stochastic resetting},\
  }\href@noop {} {\bibfield  {journal} {\bibinfo  {journal} {Physics}\ }\textbf
  {\bibinfo {volume} {1}},\ \bibinfo {pages} {40} (\bibinfo {year}
  {2019})}\BibitemShut {NoStop}%
\bibitem [{\citenamefont {Ku{\'s}mierz}\ and\ \citenamefont
  {Gudowska-Nowak}(2019)}]{kusmierz2019subdiffusive}%
  \BibitemOpen
  \bibfield  {author} {\bibinfo {author} {\bibfnamefont {{\L}.}~\bibnamefont
  {Ku{\'s}mierz}}\ and\ \bibinfo {author} {\bibfnamefont {E.}~\bibnamefont
  {Gudowska-Nowak}},\ }\bibfield  {title} {\bibinfo {title} {Subdiffusive
  continuous-time random walks with stochastic resetting},\ }\href@noop {}
  {\bibfield  {journal} {\bibinfo  {journal} {Phys. Rev. E}\ }\textbf {\bibinfo
  {volume} {99}},\ \bibinfo {pages} {052116} (\bibinfo {year}
  {2019})}\BibitemShut {NoStop}%
\bibitem [{\citenamefont {Tateishi}\ \emph {et~al.}(2020)\citenamefont
  {Tateishi}, \citenamefont {Ribeiro}, \citenamefont {Sandev}, \citenamefont
  {Petreska},\ and\ \citenamefont {Lenzi}}]{PhysRevE.101.022135}%
  \BibitemOpen
  \bibfield  {author} {\bibinfo {author} {\bibfnamefont {A.~A.}\ \bibnamefont
  {Tateishi}}, \bibinfo {author} {\bibfnamefont {H.~V.}\ \bibnamefont
  {Ribeiro}}, \bibinfo {author} {\bibfnamefont {T.}~\bibnamefont {Sandev}},
  \bibinfo {author} {\bibfnamefont {I.}~\bibnamefont {Petreska}},\ and\
  \bibinfo {author} {\bibfnamefont {E.~K.}\ \bibnamefont {Lenzi}},\ }\bibfield
  {title} {\bibinfo {title} {Quenched and annealed disorder mechanisms in comb
  models with fractional operators},\ }\href
  {https://doi.org/10.1103/PhysRevE.101.022135} {\bibfield  {journal} {\bibinfo
   {journal} {Phys. Rev. E}\ }\textbf {\bibinfo {volume} {101}},\ \bibinfo
  {pages} {022135} (\bibinfo {year} {2020})}\BibitemShut {NoStop}%
\bibitem [{\citenamefont {Ray}(2020)}]{ray2020space}%
  \BibitemOpen
  \bibfield  {author} {\bibinfo {author} {\bibfnamefont {S.}~\bibnamefont
  {Ray}},\ }\bibfield  {title} {\bibinfo {title} {Space-dependent diffusion
  with stochastic resetting: {A} first-passage study},\ }\href@noop {}
  {\bibfield  {journal} {\bibinfo  {journal} {J. Chem. Phys.}\ }\textbf
  {\bibinfo {volume} {153}},\ \bibinfo {pages} {234904} (\bibinfo {year}
  {2020})}\BibitemShut {NoStop}%
\bibitem [{\citenamefont {O'Shaughnessy}\ and\ \citenamefont
  {Procaccia}(1985{\natexlab{a}})}]{PhysRevLett.54.455}%
  \BibitemOpen
  \bibfield  {author} {\bibinfo {author} {\bibfnamefont {B.}~\bibnamefont
  {O'Shaughnessy}}\ and\ \bibinfo {author} {\bibfnamefont {I.}~\bibnamefont
  {Procaccia}},\ }\bibfield  {title} {\bibinfo {title} {Analytical solutions
  for diffusion on fractal objects},\ }\href@noop {} {\bibfield  {journal}
  {\bibinfo  {journal} {Phys. Rev. Lett.}\ }\textbf {\bibinfo {volume} {54}},\
  \bibinfo {pages} {455} (\bibinfo {year} {1985}{\natexlab{a}})}\BibitemShut
  {NoStop}%
\bibitem [{\citenamefont {O'Shaughnessy}\ and\ \citenamefont
  {Procaccia}(1985{\natexlab{b}})}]{PhysRevA.32.3073}%
  \BibitemOpen
  \bibfield  {author} {\bibinfo {author} {\bibfnamefont {B.}~\bibnamefont
  {O'Shaughnessy}}\ and\ \bibinfo {author} {\bibfnamefont {I.}~\bibnamefont
  {Procaccia}},\ }\bibfield  {title} {\bibinfo {title} {Diffusion on
  fractals},\ }\href@noop {} {\bibfield  {journal} {\bibinfo  {journal} {Phys.
  Rev. A}\ }\textbf {\bibinfo {volume} {32}},\ \bibinfo {pages} {3073}
  (\bibinfo {year} {1985}{\natexlab{b}})}\BibitemShut {NoStop}%
\bibitem [{\citenamefont {Sokolov}(1999)}]{PhysRevE.60.5528}%
  \BibitemOpen
  \bibfield  {author} {\bibinfo {author} {\bibfnamefont {I.~M.}\ \bibnamefont
  {Sokolov}},\ }\bibfield  {title} {\bibinfo {title} {Two-particle dispersion
  by correlated random velocity fields},\ }\href@noop {} {\bibfield  {journal}
  {\bibinfo  {journal} {Phys. Rev. E}\ }\textbf {\bibinfo {volume} {60}},\
  \bibinfo {pages} {5528} (\bibinfo {year} {1999})}\BibitemShut {NoStop}%
\bibitem [{\citenamefont {Richardson}(1926)}]{richardson1926atmospheric}%
  \BibitemOpen
  \bibfield  {author} {\bibinfo {author} {\bibfnamefont {L.~F.}\ \bibnamefont
  {Richardson}},\ }\bibfield  {title} {\bibinfo {title} {Atmospheric diffusion
  shown on a distance-neighbour graph},\ }\href@noop {} {\bibfield  {journal}
  {\bibinfo  {journal} {Proc. Math. Phys. Eng. Sci.}\ }\textbf {\bibinfo
  {volume} {110}},\ \bibinfo {pages} {709} (\bibinfo {year}
  {1926})}\BibitemShut {NoStop}%
\bibitem [{\citenamefont {Monin}\ \emph {et~al.}(1971)\citenamefont {Monin},
  \citenamefont {Yaglom},\ and\ \citenamefont {Lumley}}]{book1371194}%
  \BibitemOpen
  \bibfield  {author} {\bibinfo {author} {\bibfnamefont {A.~S.}\ \bibnamefont
  {Monin}}, \bibinfo {author} {\bibfnamefont {A.~M.}\ \bibnamefont {Yaglom}},\
  and\ \bibinfo {author} {\bibfnamefont {J.~L.}\ \bibnamefont {Lumley}},\
  }\href@noop {} {\emph {\bibinfo {title} {Statistical Fluid Mechanics:
  Mechanics of Turbulence, Vol. 1}}}\ (\bibinfo  {publisher} {The MIT Press},\
  \bibinfo {year} {1971})\BibitemShut {NoStop}%
\bibitem [{\citenamefont {Monin}\ \emph {et~al.}(1975)\citenamefont {Monin},
  \citenamefont {Yaglom},\ and\ \citenamefont {Lumley}}]{book1371197}%
  \BibitemOpen
  \bibfield  {author} {\bibinfo {author} {\bibfnamefont {A.~S.}\ \bibnamefont
  {Monin}}, \bibinfo {author} {\bibfnamefont {A.~M.}\ \bibnamefont {Yaglom}},\
  and\ \bibinfo {author} {\bibfnamefont {J.~L.}\ \bibnamefont {Lumley}},\
  }\href@noop {} {\emph {\bibinfo {title} {Statistical Fluid Mechanics:
  Mechanics of Turbulence, Vol. 2}}}\ (\bibinfo  {publisher} {The MIT Press},\
  \bibinfo {year} {1975})\BibitemShut {NoStop}%
\bibitem [{\citenamefont {Boffetta}\ and\ \citenamefont
  {Sokolov}(2002)}]{PhysRevLett.88.094501}%
  \BibitemOpen
  \bibfield  {author} {\bibinfo {author} {\bibfnamefont {G.}~\bibnamefont
  {Boffetta}}\ and\ \bibinfo {author} {\bibfnamefont {I.~M.}\ \bibnamefont
  {Sokolov}},\ }\bibfield  {title} {\bibinfo {title} {Relative dispersion in
  fully developed turbulence: The {Richardson's} law and intermittency
  corrections},\ }\href {https://doi.org/10.1103/PhysRevLett.88.094501}
  {\bibfield  {journal} {\bibinfo  {journal} {Phys. Rev. Lett.}\ }\textbf
  {\bibinfo {volume} {88}},\ \bibinfo {pages} {094501} (\bibinfo {year}
  {2002})}\BibitemShut {NoStop}%
\bibitem [{\citenamefont {Daniel~ben Avraham}(2000)}]{book321195}%
  \BibitemOpen
  \bibfield  {author} {\bibinfo {author} {\bibfnamefont {S.~H.}\ \bibnamefont
  {Daniel~ben Avraham}},\ }\href@noop {} {\emph {\bibinfo {title} {Diffusion
  and reactions in fractals and disordered systems}}}\ (\bibinfo  {publisher}
  {CUP},\ \bibinfo {year} {2000})\BibitemShut {NoStop}%
\bibitem [{\citenamefont {Brault}\ \emph {et~al.}(2009)\citenamefont {Brault},
  \citenamefont {Josserand}, \citenamefont {Bauchire}, \citenamefont
  {Caillard}, \citenamefont {Charles},\ and\ \citenamefont
  {Boswell}}]{PhysRevLett.102.045901}%
  \BibitemOpen
  \bibfield  {author} {\bibinfo {author} {\bibfnamefont {P.}~\bibnamefont
  {Brault}}, \bibinfo {author} {\bibfnamefont {C.}~\bibnamefont {Josserand}},
  \bibinfo {author} {\bibfnamefont {J.-M.}\ \bibnamefont {Bauchire}}, \bibinfo
  {author} {\bibfnamefont {A.}~\bibnamefont {Caillard}}, \bibinfo {author}
  {\bibfnamefont {C.}~\bibnamefont {Charles}},\ and\ \bibinfo {author}
  {\bibfnamefont {R.~W.}\ \bibnamefont {Boswell}},\ }\bibfield  {title}
  {\bibinfo {title} {Anomalous diffusion mediated by atom deposition into a
  porous substrate},\ }\href {https://doi.org/10.1103/PhysRevLett.102.045901}
  {\bibfield  {journal} {\bibinfo  {journal} {Phys. Rev. Lett.}\ }\textbf
  {\bibinfo {volume} {102}},\ \bibinfo {pages} {045901} (\bibinfo {year}
  {2009})}\BibitemShut {NoStop}%
\bibitem [{\citenamefont {Su}\ \emph {et~al.}(2005)\citenamefont {Su},
  \citenamefont {Sander}, \citenamefont {Liu}, \citenamefont {Anh},\ and\
  \citenamefont {Barry}}]{SU2005852}%
  \BibitemOpen
  \bibfield  {author} {\bibinfo {author} {\bibfnamefont {N.}~\bibnamefont
  {Su}}, \bibinfo {author} {\bibfnamefont {G.}~\bibnamefont {Sander}}, \bibinfo
  {author} {\bibfnamefont {F.}~\bibnamefont {Liu}}, \bibinfo {author}
  {\bibfnamefont {V.}~\bibnamefont {Anh}},\ and\ \bibinfo {author}
  {\bibfnamefont {D.}~\bibnamefont {Barry}},\ }\bibfield  {title} {\bibinfo
  {title} {Similarity solutions for solute transport in fractal porous media
  using a time- and scale-dependent dispersivity},\ }\href@noop {} {\bibfield
  {journal} {\bibinfo  {journal} {App. Math. Model.}\ }\textbf {\bibinfo
  {volume} {29}},\ \bibinfo {pages} {852} (\bibinfo {year} {2005})}\BibitemShut
  {NoStop}%
\bibitem [{\citenamefont {Wyld}(1999)}]{book9895}%
  \BibitemOpen
  \bibfield  {author} {\bibinfo {author} {\bibfnamefont {H.~W.}\ \bibnamefont
  {Wyld}},\ }\href@noop {} {\emph {\bibinfo {title} {Mathematical Methods for
  Physics}}},\ \bibinfo {edition} {2nd}\ ed.,\ Advanced book classics\
  (\bibinfo  {publisher} {Advanced Book Program, Perseus Books},\ \bibinfo
  {year} {1999})\BibitemShut {NoStop}%
\bibitem [{\citenamefont {Gardiner}(1996)}]{book12395}%
  \BibitemOpen
  \bibfield  {author} {\bibinfo {author} {\bibfnamefont {C.~W.}\ \bibnamefont
  {Gardiner}},\ }\href@noop {} {\emph {\bibinfo {title} {Handbook of Stochastic
  Methods}}},\ \bibinfo {edition} {2nd}\ ed.,\ Springer Series in Synergetics\
  (\bibinfo  {publisher} {Springer},\ \bibinfo {year} {1996})\BibitemShut
  {NoStop}%
\end{thebibliography}%

\end{document}